\documentclass[twocolumn,aps,prd,amsmath,amssymb]{revtex4}
\usepackage{graphicx}
\input epsf

\bibpunct{}{}{,}{s}{}{,}
\usepackage{graphicx}   
\begin{document}
\title{Fuzzballs and the information paradox: a summary and conjectures}

\author{Samir D. Mathur}
\email{mathur@mps.ohio-state.edu}
\affiliation{The Ohio State University, Columbus, OH 43210, USA}

\begin{abstract}

 The black hole information paradox is one of the most important issues in theoretical physics. We review some recent progress  using string theory in understanding the nature of black hole microstates. For all cases where these microstates have been constructed, one finds that they are horizon sized `fuzzballs'. Most computations are for extremal states, but recently one has been able to study a special family of non-extremal microstates, and see `information carrying  radiation' emerge from these gravity solutions. We discuss how the fuzzball picture can resolve  the information paradox. We use the nature of fuzzball states to make some conjectures on the dynamical aspects of black holes, observing that the large phase space of fuzzball solutions can make the black hole   more `quantum' than assumed in traditional treatments.

\bigskip

{\em Black holes, String theory}: 
\end{abstract}

\def\b{\bigskip}
\def\be{\begin{equation}}
\def\bea{\begin{eqnarray}}
\def\ee{\end{equation}}
\def\eea{\end{eqnarray}}
\def\r{\rightarrow}
\def\p{\partial}
\def\t{\tilde}
\def\nn{\nonumber\\ }
\def\h{{1\over 2}}

\maketitle

\section{The information paradox}\label{introduction}

Most people have heard of the black hole information paradox \cite{hawking}. But the full strength of this paradox is not always appreciated. If we make two reasonable sounding assumptions

\b

(a) All quantum gravity effects  die off rapidly at distances beyond some fixed length scale (e.g. planck length $l_p$ or string length $l_s$)

\b

(b) The vacuum of the theory is unique

\b

Then we {\it will} have `information loss' when a black hole forms and evaporates, and quantum unitarity will be violated. (The Hawking `theorem' can be exhibited in this form \cite{exactly}, and it can be seen from the derivation how conditions (a),(b) above can be made more precise  and the `theorem' made as rigorous as we wish.)

In this article we will see that string theory gives us a way out of the information paradox, by violating assumption (a). How can this happen? One usually thinks that the natural length scale for quantum gravity effects is $l_p$, since this is the only length scale that we can make from the fundamental constants $G, \hbar, c$. But a black hole is a large object, made by putting together some large number $N$ 
of fundamental quanta. Thus we need to ask whether  non-classical effects
extend over distances $l_p$ or over distances $N^\alpha l_p$ for some constant $\alpha$. One finds that the latter is true, and that the emerging length scale for quantum corrections is order horizon radius. The information of the hole is distributed throughout a horizon sized `fuzzball'. Hawking radiation is thus  not emitted from a region which is an `information free vacuum', and the information paradox is resolved.

\subsection{Emission from the black hole}

To see how the information paradox arises, we must first see how Hawking radiation is produced in the traditional picture of the black hole. Consider the semiclassical approximation, where we have a quantum field living on a classical spacetime geometry.  If the metric of this spacetime is time dependent, then the quantum field will not in general sit in a given vacuum state, and pairs of particles will be produced. The Schwarzschild black hole has a metric
\be
ds^2=-(1-{2M\over r}) dt^2+{dr^2\over 1-{2M\over r}}+r^2d\Omega_2^2
\ee
This metric looks time independent, but that is an illusion; these Schwarzschild coordinates cover only the exterior of the hole, and if we look at the full geometry of the spacetime then we cannot obtain a time independent slicing of the geometry. 

\begin{figure}[ht]
\includegraphics[scale=.15]{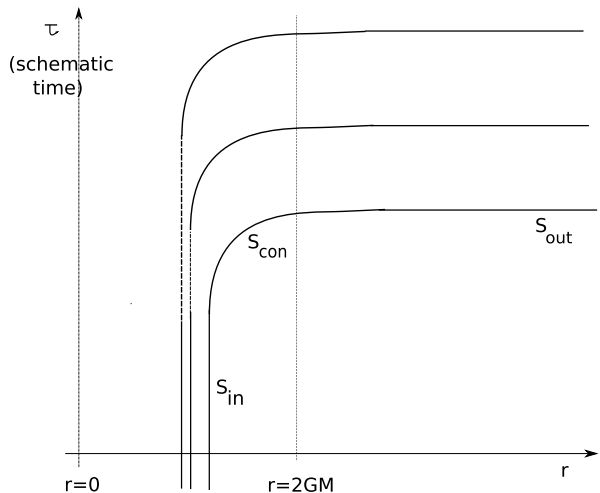}
%
%
\caption{Constructing a slicing of the black hole geometry. For $r>3GM$ we have the part $S_{out}$ as a $t=constant$ slice. The `connector' part $S_{con}$ is almost the same on all slices, and has a smooth intrinsic metric as the surface crosses the horizon. The inner part of the slice $S_{in}$ is a $r=constant$ surface, with the value of $r$ kept away from the singularity at $r=0$. The coordinate $\tau$ is only schematic; it will degenerate at the horizon. }
\label{matfthree}       
\end{figure}

We schematically sketch some spacelike slices for the Schwarzschild geometry in fig.\ref{matfthree}.  (This figure is not a Penrose diagram; it is just a formal depiction of the exterior and interior regions of the hole, and if we try to put any time independent coordinates on this space they will degenerate at the horizon $r=2M$.) Outside the horizon ($r>2M$) we can take the spacelike slice to be $t=t_0$; this part is called $S_{out}$ in the figure. Inside the horizon $(r<2M)$  the constant $t$ surface is {\it timelike}. We get a spacelike surface by taking  $r=r_0$ instead; this part is termed $S_{in}$.  We can join these two parts of the spacelike surface by a `connector region' $S_{con}$, so that we construct a spacelike surface  covering regions both  outside and  inside the horizon.  The details of such a construction can  be found in the reference listed above \cite{exactly}, and we will summarize the discussion given there.

 How do we make a `later' spacelike slice? Outside the horizon we can take the surface $t=t_0+\Delta t$. Inside the horizon we must now continue our constant $r$ surface for a little longer before joining it to the constant $t$ part. Thus the later surface is not identical in its intrinsic geometry to the earlier one.  We have a time dependent slicing, and there will be particle production in the region where the surface is being `stretched'. 

\begin{figure}[ht]
\includegraphics[scale=.15]{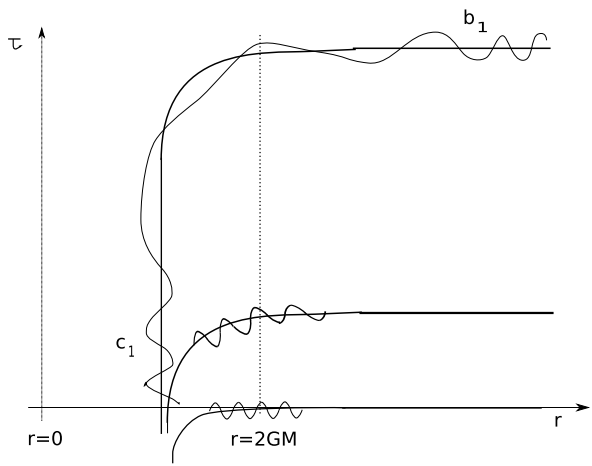}
%
%
\caption{A fourier mode on the initial spacelike surface is evolved  to later spacelike surfaces. In the initial part of the evolution the wavelength increases but there is no significant distortion of the general shape of the mode. At this stage the initial vacuum state is still a vacuum state. Further evolution leads to a distorted waveform, which results in particle creation.}
\label{matffourtp}       
\end{figure}

To see this particle production consider the evolution of wavemodes in the geometry. To leading order we can evolve the wavemode by letting the surfaces of constant phase lie along the null geodesics of the geometry. Fig. \ref{matffourtp} shows a wavemode being stretched and deformed, so that even though the wavemode was not populated by particles at the start of the evolution, we have some amplitude to get particles $b_1$ and $c_1$ at the end of the stretching. The crucial point here is the state of these created quanta. This state has the form   $e^{\gamma b_1^\dagger c_1^\dagger}|0\rangle$, where $\hat b_1^\dagger$ creates quanta on the part of the slice outside the horizon and $\hat c_1^\dagger $ creates quanta on the part of the slice inside the horizon. This state  can thus be expanded in a series of terms that have  $0, 1, 2 \dots$ particle pairs. To understand the essentials of the paradox we can replace the state by a simpler one with just two terms
\be
|\psi\rangle_1={1\over \sqrt{2}}[~|0\rangle_{b_1}\otimes|0\rangle_{c_1}+ |1\rangle_{b_1}\otimes|1\rangle_{c_1}~]
\label{two}
\ee
We see that the state of quanta outside the horizon (the $b$ quanta) is `entangled' with the state of the quanta inside the horizon (the $c$ quanta). 

\begin{figure}[ht]
\includegraphics[scale=.15]{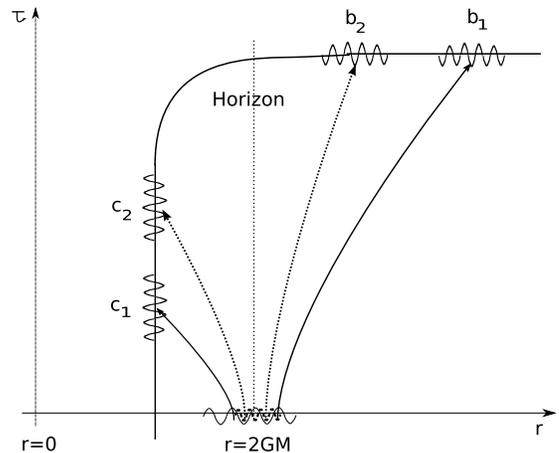}
%
%
\caption{On the initial spacelike slice we have depicted two fourier modes: the longer wavelength mode is drawn with a solid line and the shorter wavelength mode is drawn with a dotted line. The mode with longer wavelength distorts to a nonuniform shape first, and creates an entangled pairs $b_1, c_1$. The mode with shorter wavelength evolves for some more time before suffering the same distortion, and then it creates  entangled pairs $b_2, c_2$.}
\label{matftthree}       
\end{figure}

 It is important to see how the next pair of quanta are created (fig.\ref{matftthree}). The spacelike slice stretches, moving the locations of the $b_1, c_1$ quanta further apart. In the new region that is created, an entangled pair $b_2, c_2$ is created out of the vacuum. Thus the overall state can be written schematically in the form
 \be
 |\psi\rangle=\prod_k {1\over \sqrt{2}}~[ |0\rangle_{b_k}\otimes|0\rangle_{c_k}+ |1\rangle_{b_k}\otimes|1\rangle_{c_k}]
 \label{one}
\ee

\subsection{The problem with the entangled state}

To see how the above state leads to the information paradox, let us make some basic observations.

\b

(i) The state $|\psi\rangle$ is `highly entangled' between the $b, c$ pairs. We can compute  the entropy of this entanglement by tracing over the $c$ quanta, obtaining the density matrix $\rho$ describing the $b$ quanta, and computing the entropy $S=-{\rm Tr}[ \rho\log \rho]$ of this density matrix. This entropy is of order the Bekenstein entropy \cite{bek} of the hole. If the hole evaporates away  completely then we are left with the $b$ quanta in their highly  entangled state but we cannot see anything that they are entangled {\it with}. Thus an initial pure state which formed the hole has evolved to a mixed state, and we have lost unitarity.

\b

(ii) A common misconception is that `subtle quantum gravity effects' can change the state of the emitted radiation and  resolve this problem. This is incorrect. Suppose we change the state of each entangled pair in (\ref{one}) a little,
 \be
 |\psi'\rangle=\prod_k {1\over \sqrt{2}}[(1+\epsilon_k) |0\rangle_{b_k}\otimes|0\rangle_{c_k}+ (1-\epsilon_k)|1\rangle_{b_k}\otimes|1\rangle_{c_k}]
 \label{onep}
\ee
where $|\epsilon_k|\ll 1$.  Then the state is still highly entangled; the entropy of entanglement has changed by a very small fraction. A pure state for the $b$ quanta would be a state like 
\be
 |\psi''\rangle=\prod_k [{1\over \sqrt{2}} (  |0\rangle_{b_k}+|1\rangle_{b_k})]\otimes  [{1\over \sqrt{2}} (  |0\rangle_{c_k}+|1\rangle_{c_k})]
 \label{oneq}
\ee
But such a state is nowhere `close' to the state (\ref{one}); we need an {\it order unity} change in the state of each pair $b_k, c_k$. 

\b

\begin{figure}[ht]
\includegraphics[scale=.20]{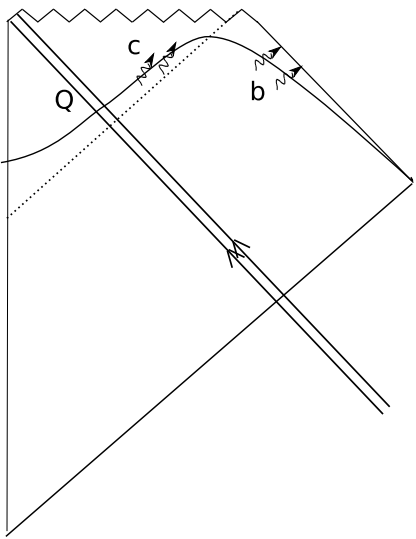}
%
%
\caption{ The infalling matter $Q$ and the entangled pairs $c,b$ shown on the spacelike slices in the Penrose diagram. }
\label{matfseventq}       
\end{figure}

(iii) If we somehow obtained a state like (\ref{oneq}) then the emitted radiation would be in a pure state, but this would still not help; the state of the radiation would have  no dependence on the initial matter making the hole. Fig.\ref{matfseventq} shows a Penrose diagram of the hole. On any spatial slice there are   three kinds of matter that we must consider.  On the extreme left we have the infalling matter $Q$ that made the hole.  Next we have the `negative energy quanta' $c_k$ and finally near spatial infinity we have the quanta $b_k$. What we need is for the $b_k$ to form a pure state (entangled with nothing else), but carrying the information of the initial matter $Q$. 

\b

(iv) So what prevents the information of $Q$ from reaching the quanta $b_k$? When we burn a piece of coal, the emitted radiation does manage to carry all the information of the coal. The first quantum emitted from the coal may well be in a mixed state with the part of the coal left behind; for example the emitted quantum may be a photon, and its spin may be entangled with the spin of the emitting atom which stays behind in the coal
\be
|\chi\rangle={1\over \sqrt{2}}[ ~|\uparrow\rangle_{\rm photon}\otimes|\downarrow\rangle_{\rm atom}+|\downarrow\rangle_{\rm photon}\otimes|\uparrow\rangle_{\rm atom}~
]
\ee
The next quantum emitted from the coal may also be in a mixed state with the coal, but note that the emission process will be influenced in principle  by the spin of the atom left behind after the first emission. In this way the spin of later emitted quanta get related to the spins of earlier emitted quanta, and if the coal finally burns away to nothing then the emitted radiation survives in a pure state, with all the information of the initial piece of coal.

We can now see the difference between this process and the evaporation of the hole. The radiation quanta $b_k, c_k$ are pairs created from the {\it vacuum}. The matter $Q$ is far away (several miles for a typical astrophysical hole) from the place where the spacelike slice is stretching and producing quanta, so its information does not influence the state of the created pairs.  Further, later pairs $b_k, c_k$ are produced in a way that does not depend on the state of earlier produced pairs.   As we had seen  from fig.\ref{matftthree},  after the quanta $b_1, c_1$ are created, the part of the spacelike slice carrying these quanta stretches in such a way that these quanta are moved away from the region near the horizon where the production of the next pair $b_2, c_2$ will occur. Thus unlike the case of the coal, here the the state of later pairs does not depend on the state of earlier pairs. 

\b

(v) We can now summarize the essential strength of the information paradox. The region near the horizon has a curvature length scale $\sim M$, which we can take to be  of order several miles. Consider the evolution of modes of a quantum field in this region. Follow the evolution of a field mode from the time its wavelength is say $M/100$ to the time it stretches to a wavelength $\sim M$; this evolution takes a `time' $\sim M$. With all length and time scales being classical, and the evolution taking place far away from the matter $Q$ and any region of high curvature, the evolution of the mode will lead to a state like (\ref{two}). But to solve the information problem {\it we need the actual evolution of the field mode in this situation  to differ by order unity from the expected evolution}. 

\section{The fuzzball program}

\begin{figure}[ht]
\centerline{\epsfxsize=2.5in\epsfbox{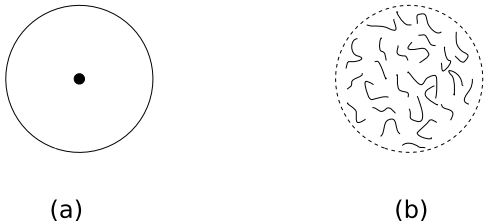}}   
\caption{(a) The conventional picture of a black hole (b) the proposed picture -- information of the state is distributed throughout the `fuzzball'. \label{inter}}
\end{figure}

The fuzzball program solves the paradox by showing that assumption (a) in the above section is incorrect; quantum effects change the black hole interior in a way that distributes the information of the hole throughout a horizon sized region.  Schematically, the  picture of the hole is changed from that in fig.\ref{inter}(a) to that in fig.\ref{inter}(b), where the latter picture shows a `quantum fuzz' filling a horizon sized region. The modification of the black hole interior allows the emitted quanta to carry the information of the state of the hole. 

\begin{figure}[ht]
\includegraphics[scale=.12]{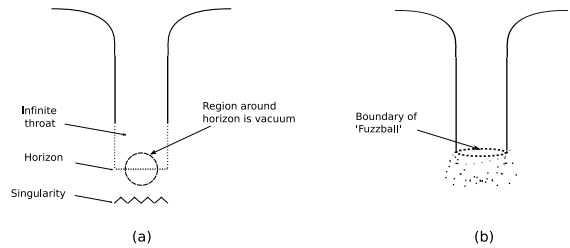}
%
%
\caption{(a) The traditional geometry of the extremal hole; the state near the horizon is the {\it vacuum} with no information about the microstate (b) The fuzzball proposal; there is no `information free horizon' region like the sphere sketched in (a).  }
\label{firstp}       
\end{figure}

While astrophysical holes are typically charge neutral, in string theory it is easier to start with supersymmetric holes which have a charge equal to their mass. Thus they are `extremal black holes' in general relativity, and give supersymmetric solutions in string theory. The traditional picture of the extremal hole is shown schematically in fig.\ref{firstp}(a). We have flat space at infinity, then a `neck' leading to an infinite `throat'. There is a horizon at the end of the throat, through which a quantum can fall in finite {\it proper} time. There is a region inside the horizon, which contains a timelike singularity. The  region around the horizon is a low curvature region. The important point is that  if we draw a ball shaped region around the horizon then the state in this region is the  {\it vacuum}  state $|0\rangle$. Thus there is no information about the hole in the vicinity of this horizon.  We will term  a horizon like this with no information in its vicinity an `information free horizon'. 

Fig.\ref{firstp}(b) depicts the extremal hole given by the fuzzball proposal. We have flat space at infinity, the neck and the throat. But while the throat is long, it is not infinite. The throat ends in a quantum fuzzy `cap', where the precise details of the cap contain the information of the state of the hole. 

\subsection{The fuzzball proposal}

The fuzzball program is primarily a {\it construction.} We take a specific black hole in string theory, with some mass and charges. This hole should have $e^{S_{bek}}$ microstates, where $S_{bek}$ is the Bekenstein entropy of the hole. We try to construct these microstates  and see what they look like. All cases worked out so far have given microstates that are `fuzzballs'; there is no horizon, and the details of  the microstates are explicitly manifested by the gravity solution. In particular, all extremal black hole states that have been constructed have the form  fig.\ref{firstp}(b), and not the form fig.\ref{firstp}(a).\footnote{The motivation for expecting that black hole microstates are fuzzballs comes from an estimate about the size of string theory bound states \cite{emission} (the computations are reproduced in more detail in one of the reviews mentioned above \cite{review1}). One finds that when different charges in string theory bind together, `fractional' excitations develop. These fractional excitations are very low tension, floppy objects, which stretch far, and make the size of a bound state the same order as the traditional horizon radius.} 

Note that if fig.\ref{inter}(b) or fig.\ref{firstp}(b) was the true picture of all black hole microstates then there would be no information paradox. An infalling quantum would not encounter a vacuum all the way to a singularity,   but instead would interact with the degrees of freedom of the hole, just like what happens when a photon falls on a piece of coal. 

So far we have a good understanding of all states for the 2-charge extremal hole (the so called `small black hole'), and we also understand large sets of microstates for the 3-charge and 4-charge extremal holes. One family of states for the non-extremal hole has also been constructed; moreover, these nonextremal states are found to emit radiation at exactly the rate that would be expected for the `Hawking emission' from these special microstates. (For some reviews on the fuzzball program, see \cite{review1, review2, warner, skenderisr}.)

The fuzzball `conjecture' says that all microstates of all black holes will behave like the ones that have been constructed.  Let us see in more detail what this means. The essential property of the microstates found in the fuzzball program is that there is no `information free horizon'. Consider first the extremal hole. In the traditional picture fig.\ref{firstp}(a) we can mark a ball shaped region around the horizon  where all quantum fields are in the {\it vacuum} state $|0\rangle$; i.e., we just have the expected vacuum of quantum fields on gently curved spacetime. With the fuzzball conjecture {\it it is not possible to find such a ball shaped region around a horizon}. While the redshift may be large near the fuzzy region drawn in fig.\ref{firstp}(b), there is no region that we can mark out that will look like a piece of the traditional extremal Penrose diagram straddling the horizon. Any ball shaped region we draw near the fuzzball boundary will have a state $|\psi\rangle$ that is {\it not} near the vacuum state $|0\rangle$. Rather, we will have 
\be
\langle 0 | \psi\rangle \rightarrow 0 ~~~{\rm for} ~~~{M\over m_{pl}}\rightarrow \infty
\ee
so that the state $|\psi\rangle$ would be nearly orthogonal to the vacuum $|0\rangle$ for holes with large mass $M$. 

This absence of a traditional horizon distinguishes the fuzzball proposal from many other attempts to understand the information problem. Let us list some of these alternative proposals. First, we have Hawking's original proposal which says that information is indeed lost, and we should build our quantum theory without requiring a unitary S-matrix. Another proposal  is that the information moves into baby Universes forming inside the horizon region. Another recent proposal is that we should impose a `final state boundary condition' at the black hole singularity \cite{hormalda},  so that information is forced to   emerge in the Hawking radiation. By contrast, the fuzzball proposal  does not require `new physics'. Instead the proposal says that when we actually construct the microstates of a black hole in the full theory of quantum gravity then we find the state to be a `puffed up fuzzball', and so radiation from the microstate is no different from radiation from a piece of coal. 

Before proceeding to see in more detail what kind of microstates we find for black holes, let us note some common misconceptions about the information puzzle and the fuzzball proposal. 

\bigskip

(a)  AdS/CFT duality is one of the most remarkable results to emerge from string theory \cite{maldacena}. It is sometimes believed that we can resolve the information paradox by using this duality. This is incorrect, since such an argument would be circular. As we discussed in the last section, if we are given assumptions (a),(b) about quantum gravity then we {\it will} have a breakdown of quantum unitarity. In this situation we will also lose the AdS/CFT correspondence, since this duality assumes that both sides of the duality are good unitary quantum theories. Thus to save quantum theory (and AdS/CFT in particular) we have to show that at least one of the traditional assumptions (a),(b) breaks down in our full theory of quantum gravity. We have to resolve the problem in the {\it gravity} description of the state; it is a circular argument to say that information will come out because there is a dual field theory that is unitary.

This said, it will turn out that the AdS/CFT correspondence will be a very important tool in helping us understand the general set of microstates. It is easier to count and classify states in the CFT, so while we must construct our microstates in the gravity picture to resolve the information paradox, we can use the CFT analysis to know when we have constructed all the states (or enough that the general state can be understood as an extrapolation of those that have been made).

\bigskip

(b) A common question about fuzzballs is: does an infalling observer feel something very different when falling into a fuzzball than into a traditional black hole? This is a dynamical question, and we will try to use our knowledge of the time independent fuzzball states to conjecture an answer in section \ref{dynamical}. The key point will be that there are different energy and time scales for different processes. For {\it heavy} observers (mass much larger than the Hawking temperature) and over short times (order the infall time) the behavior of the typical fuzzball may be no different from the behavior of the traditional black hole geometry. But over {\it long} times (order the Hawking evaporation time) the fuzzball behaves differently from the traditional black hole, and returns information to infinity in the Hawking radiation while the traditional black hole geometry leads to information loss.

There are a couple of things that we need to be careful about when addressing such issues. First, it is sometimes believed that if the fuzzball state is `too complicated' then it is `essentially' the vacuum, and should be replaced by $|0\rangle$. This is incorrect. The generic fuzzball state is indeed very `complicated', but it is important that it is close to being {\it orthogonal} to the vacuum. All we can say is that for some {\it particular} process the fuzzball state behaves almost like the vacuum state. Secondly, it is sometimes believed that the fuzzball state will have a `fine structure' that will affect only motion over planck distances; evolution of ordinary quanta will be just the same as in the traditional black hole geometry. This is incorrect; in fact as noted in section \ref{introduction} (and shown in detail in the reference mentioned above \cite{exactly}) we need the evolution of Hawking wavelength quanta to change {\it by order unity} at the horizon. We will note below that for the one family of non-extremal microstates that are known, the low energy emitted quanta indeed see the detailed structure of the `ergoregion' of the geometry, while high energy quanta are not sensitive to the location of the ergoregion.

\b

Before moving to a detailed study of fuzzballs, let us ask 
 what would constitute a `disproof' of the fuzzball conjecture.  To disprove the conjecture we would have to show that generic states of the hole {\it do} have an `information free horizon'. For extremal holes, this would need us to argue that there are two kinds of microstates: the ones that are like the `fuzzballs' that have so far been found, and the remainder that are {\it not} like fuzzballs. With all we know now this looks hard to do, since in the dual CFT description there seems to be no sharp boundary between different classes of microstates, and for the simple case of the 2-charge extremal hole {\it all} states have been found to be fuzzballs.

\section{Black holes in string theory}

The remarkable thing about string theory is that it admits no free parameters -- it is a unique theory with all brane tensions and couplings fixed. There is of course a large freedom in which solution of the theory we choose to look at; this freedom  allows us for example to choose any value for the dilaton field which sets the local value of the string coupling $g$.

Since we cannot add anything to the theory, we must make our black hole from objects {\it in} the theory.  The theory contains gravitons, as any theory of quantum gravity would, and a collection of extended objects - strings and branes - of different dimensionalities. One knows that all different versions of string theory are related by exact dualities, so we can use any one; we will take type IIB string theory for concreteness. 

One makes black holes by taking branes in the theory and wrapping them on compact directions; from the viewpoint of the noncompact directions this places a given mass at a point in space, and with a suitable choice of wrapped objects we can create a black hole. 
Among the objects in IIB string theory we have 5-dimensional branes, which we will use. Thus we compactify five directions as follows
\be
M_{9,1}\r M_{4,1}\times S^1\times T^4
\label{sfour}
\ee
where we have singled out one $S^1$ for later use.  the $S^1$ has length $2\pi R$ and the $T^4$ has volume $(2\pi)^4 V$. 

We can wrap a large number $n_1$ of strings on the $S^1$, and this does give a large mass at one point in the noncompact space. But the strings carry charge as well, and also create distortions of the moduli -- the sizes of the compact directions. When all these effects are taken into account one finds that one does not get a horizon, and there is no black hole. From a statistical entropy perspective this is good, since the degeneracy of the string bound state does not grow with $n_1$; the strings bind together by just making one `multiwound string' which loops $n_1$ times around the $S^1$ before closing on itself. Thus the statistical entropy vanishes, in agreement with the vanishing of the Bekenstein entropy.

We can do better by adding $n_p$ units of momentum along $S^1$ to the string. The strings are called the `NS1-brane' in string theory, and momentum is usually denoted `P', so this system would be called the NS1-P system. From the viewpoint of the noncompact directions, The momentum carries  Kaluza-Klein charge under the gauge field arising from reduction along $S^1$, so we have two kinds of charges in the state and this is called the 2-charge system. The extremal states of this system are those that have the minimum charge for given mass, and these turn out to be supersymmetric. What is remarkable is  that these lowest mass states are very numerous. As we had seen, the strings join up into one long string, and the momentum will bind to this string by creating travelling waves along the string. The total momentum can be partitioned among different harmonics in many ways, and each such state has the same energy. The number of states is \cite{sen}
\be
{\cal N}\sim e^{2\sqrt{2} \pi \sqrt{n_1n_p}}
\ee
so that the microscopic entropy is
\be
S_{micro}=2\sqrt{2}\pi \sqrt{n_1n_p}
\label{entropy}
\ee

\begin{figure}[ht]
\includegraphics[scale=.12]{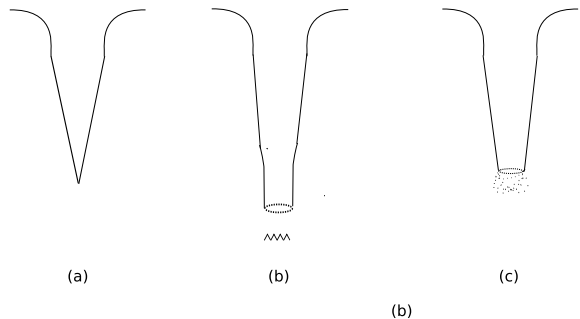}
%
%
\caption{(a) The naive NS1-P geometry assuming  the Einstein action; there is no horizon, and the metric ends in a point singularity (b) The naive geometry when we include $R^2$ terms from string theory; there is an infinite throat, ending in a horizon with a singularity inside (c) The actual geometries of the NS1-P system; the throat ends in `caps', with different caps for different microstates. The boundary of the `cap' region  shown by the dotted circle has area satisfying $A/G\sim S_{micro}$. }
\label{geometries}       
\end{figure}

What about the metric that this NS1-P system will generate? Let us discuss this in three steps:

\b

(1) At first it may seem reasonable to assume that all the strings and momentum charges sit at one location $r=0$ in the noncompact space; after all we had made a bound state of these objects and so all charges should be concentrated at a given location. This gives what we will call the `naive' geometry of the 2-charge extremal system
 \bea
ds^2&=&{1\over 1+{Q_1\over r^2}}[-dt^2+dy^2+{Q_p\over r^2}(dt+dy)^2]\nn
~~~~&+&\sum_{i=1}^4 dx_idx_i+\sum_{a=1}^4 dz_adz_a
\label{three}
\eea
Here $y$ is along $S^1$, $z_a$ are coordinates for $T^4$ and in the 4-d noncompact space $\{x_i\}$ we write $r^2=x_ix_i$. This metric has a singularity at $r=0$ but no horizon. This metric is sketched in fig.\ref{geometries}(a).

\b

(2) We note that near $r=0$ the curvature of (\ref{three}) diverges, so if there are higher derivative $\sim R^2$ terms in the gravity Lagrangian then they can be important. In string theory there is indeed a whole series of such higher derivative terms in the effective action \cite{dewit}, and it is unclear how to compute the net effect from all these terms. Dabholkar  \cite{dabholkar}  considered the 2-charge system was taken with a slightly different compactification (the $T^4$ in (\ref{sfour}) was replaced by another 4-manifold called K3). Only the first of the higher derivative corrections was considered, and it was found that the naive geometry changed to one of the kind expected for an extremal hole: there is an infinite throat ending in a horizon (fig.\ref{geometries}(b)). In the presence of higher derivative terms the Bekenstein entropy gets replaced \cite{wald} by its generalization, the Bekenstein-Wald entropy $S_{bw}$, and it was found that
\be
S_{bw}=S_{micro}
\ee
An order of magnitude agreement between these entropies had been earlier conjectured in \cite{sen}.

Thus we see that while there are still open questions about the gravity solution of the 2-charge hole, it does seem plausible that this is a simple example of an extremal black hole. The strongest argument for thinking of the 2-charge system as a good black hole comes from the form of the microscopic entropy. We have $S_{micro}\sim \sqrt{n_1n_p}$ for two charges, $S_{micro}\sim \sqrt{n_1n_pn_5}$ for three charges, and $S_{micro}\sim \sqrt{n_1n_pn_5n_k}$ for four charges; further the entropy arises in each case as a partition of momentum of a string or `effective string'. Thus let us investigate further this 2-charge system and see  how fuzzballs arise.

\b

(3) Let us now ask if (\ref{three}) is indeed the correct metric for the system. We have seen that different microstates arise from different ways of carrying the momentum on the string. The first point to note is that the fundamental string has no longitudinal vibrations, so to describe the momentum carrying wave we have to specify both the harmonic order along the string as well as the transverse direction chosen for vibration. 

\begin{figure}[ht]
\includegraphics[scale=.12]{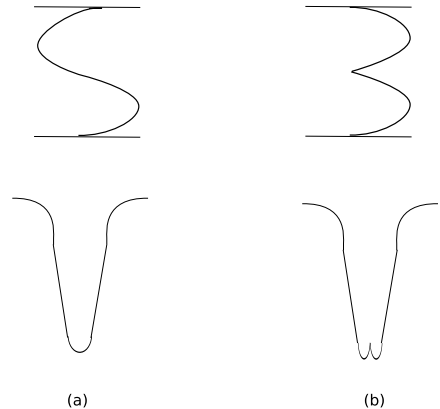}
%
%
\caption{(a) The NS1 carries the momentum P by swinging in a uniform helix with one turn in the covering space. Below, we sketch the geometry it produces; there is no horizon (b) The NS1 has no angular momentum, as the first half swings clockwise and the second anticlockwise; nevertheless, the geometry is not the naive geometry of the nonrotating NS1-P system. }
\label{strings}       
\end{figure}

To picture the vibrations of the string let us open it up to its full length $2\pi R n_1$; i.e. go to the $n_1$ fold cover of the $S^1$. 
Let us start by putting all the momentum in the lowest allowed harmonic, and choose the polarization of the vibration such that the  string in the covering space executes one turn of a uniform helix; the helix will project to a circle $x_1^2+x_2^2=a^2$ in the noncompact space. Thus the string looks like a `slinky', winding around the $S^1$ direction as it wanders around in the $x_1-x_2$ plane. The important part here is that the string is not sitting at $r=0$ in the noncompact space; instead it is spread out over a sizable region (the size of this region scales with the charges as $\sim \sqrt{n_1n_p}$). As a result  the metric produced by this vibrating string will differ from the naive expectation (\ref{three}). This metric can be  written down in a straightforward way \cite{lm3} (it was earlier found in a dual form  in related contexts \cite{cy,bal,mm}).  The metric has no horizon, and we have pictured the string and its metric in fig.\ref{strings}(a). 

What do we make of this metric? This way of choosing the momentum harmonics is certainly one of the microstates that we were counting in the entropy (\ref{entropy}). But the geometry does not agree with  (\ref{three}), and even if we apply the higher derivative corrections, we do not get the infinite throat of fig.\ref{geometries}(b). 

One might think that  the departure from (\ref{three}) arises because this particular state of the string has a large rotation; by choosing the string to swing in a helical fashion we gave the state its maximal possible angular momentum. To address this issue, let us look at a microstate that has {\it no} angular momentum. Since we know how to make all microstates, this is easy to do. Again consider our vibrating string, but let the first half of the string describe a clockwise helix, and the other half an anticlockwise helix (fig.\ref{strings}(b)). The net angular momentum will be zero. Naively, one might have thought that now we should get back the solution (\ref{three}); after all the state we are making has the same mass, charges and angular momentum as the metric (\ref{three}). But we see immediately that we will {\it not} get the metric (\ref{three}); the string has again spread over a region whose radius scales as $\sqrt{n_1n_p}$. So we see that the actual microstates of our system do not give the `naive' metric (\ref{three}). Further the region over which the metric departs from the naive metric is so large for these states that the higher derivative corrections turn out to have no significant effect on the geometry; in particular it does not change the microstate geometry to one with an infinite throat.

Let us now consider the general state of this system. Each harmonic of vibration of the string behaves like a harmonic oscillator, and strictly speaking we should specify the state of the string by giving the excitation number for each oscillator. Thus an energy eigenstate would be written like
\be
|\psi\rangle=(\hat a^{ i_1\dagger}_{k_1})^{m_1} (\hat a^{i_2\dagger}_{k_2})^{m_2}\dots (\hat a^{i_s\dagger}_{k_s})^{m_s}|0\rangle
\label{sstate}
\ee
where $|0\rangle$ is the state of the string with no vibrations, and the 
creation operator $\hat a^{i_1\dagger}_{k_1}$ creates an excitation in the harmonic $k_1$ with vibration direction $i_1$. A generic state will have $m_i\sim 1$, and so we should really write down the quantum wavefunction of the appropriate eigenstate  for each harmonic oscillator. But it is easier to start with the case where the energy of the state is placed in relatively few harmonics, so that $m_i\gg 1$. In this case we have large occupation numbers for the excited oscillators, and we can replace the energy eigenstates by coherent states without losing the essential physics of the state. Now we can describe the string by a classical vibration profile 
\be
\vec F(t-y)
\ee
where the vector $\vec F$ is transverse to the direction $y$, and is a function of only $t-y$ because the momentum moves purely upwards along the string in the extremal state (if we had vibrations going in both directions on the string then we would have more energy than needed to give the net momentum charge of the state). Let the string vibrations be in the noncompact directions. The metric of the string carrying such a vibration profile is given by \cite{lm4}
 \bea
ds^2_{string}&=&H[-dudv+Kdv^2+2A_i dx_i dv]\nn
&&~~+\sum_{i=1}^4 dx_idx_i+\sum_{a=1}^4 dz_adz_a\nn
B_{uv}&=&-{1\over 2}[H-1], ~~\qquad B_{vi}=HA_i\nn
e^{2\phi}&=&H
\label{ttsix}
\eea
where
\bea
H^{-1}&=&1+{Q_1\over L_T}\int_0^{L_T}\! {dv\over |\vec x-\vec F(v)|^2}\\
K&=&{Q_1\over
L_T}\int_0^{L_T}\! {dv (\dot
F(v))^2\over |\vec x-\vec F(v)|^2}\\
A_i&=&-{Q_1\over L_T}\int_0^{L_T}\! {dv\dot F_i(v)\over |\vec x-\vec F(v)|^2}
\label{functionsq}
\eea
where we have written $ds^2_{string}$ to denote the fact that this metric is in the `string frame' of string theory, and we have given the metric, gauge field $B$ and dilaton field $\phi$ which are the nonzero fields in this solution.

In fig.\ref{geometries}(c) we depict these solutions schematically. There is no horizon; instead the throat ends in a cap whose structure depends on the choice of profile function $\vec F$. (The same geometries can also be obtained in the language of `supertubes', where the charges are dualized to NS1-D0 \cite{supertubes}.) For string vibrations in the $T^4$ directions the metrics can be found in a similar way \cite{lmm}, and  fermionic excitations can be added  \cite{marika} to make general extremal solutions. The extremal and near-extremal behavior of 2-charge solutions have been studied in many different ways \cite{skenderis,dynamic}.  

It is now interesting to look at a generic state from the set of allowed states, and note at what radius $r\sim r_0$ this departure from the naive geometry becomes significant. Suppose we  compute the area $A$ of this surface $r=r_0$ in the naive geometry (\ref{three}); since the naive geometry and the actual geometries pretty much agree at this location what we are computing is the area of the boundary of the `fuzzball region' in a typical microstate. Interestingly, one finds that \cite{lm5}
\be
{A\over G}\sim \sqrt{n_1n_p}\sim S_{micro}
\label{ssize}
\ee
So we see that even though there is no horizon for any microstate, the boundary area of the typical microstate satisfies a Bekenstein like relation with the entropy of the system. 

\section{The D1D5 system}

In the above section we looked at extremal holes made with two charges -- NS1 and P. In string theory  we have S and T dualities, which can change one set of charges into another. These are exact symmetries of the theory, so the physics in the two descriptions will be equivalent. But it can be more convenient to describe the physics in one duality frame than in another. 

To see the structure of the 2-charge system it is useful to start with the NS1-P frame, as we did. This is because the bound states of this system are  just states of  a fundamental string carrying momentum, and it is possible to construct the metric produced by such a string. It is not obvious how to construct the metrics of the 2-charge extremal states if we start in any other duality frame. But once we have the metrics in the NS1-P frame, we can of course apply the S,T dualities to get the metrics in any other duality frame.

Why should we be interested in other duality frames? We have two goals:

\bigskip

(1) First, we would like to study small excitations around the extremal states that we have constructed. The extremal states themselves are supersymmetric ground states of system for its given charge, and they have no dynamics. If we excite the system with extra energy, then the state will become non-supersymmetric and time-dependent, and we will be able to observe the dynamical behavior of excitations around our microstate. 

\bigskip

(2) Second, we would like to construct extremal states of the extremal black hole with {\it three} charges. The 3-charge hole has a larger entropy, and therefore a larger horizon, than the 2-charge hole. The higher derivative corrections are thus small at the horizon of the 3-charge hole, so this hole looks closer to the black holes that we are familiar with. 

\bigskip

We will find it easier to do both these things if we first take our 2-charge system to another duality frame. Under the dualities the following will happen:

\bigskip

(1) The $n_1$ NS1 strings will be transformed to $n_1\equiv n'_5$ D5 branes. In the compactification  (\ref{sfour}) these D5 branes are wrapped on $T^4\times S^1$.

\bigskip

(2) The $n_p$ units of momentum will be transformed into $n_p\equiv n'_1$ D1 branes wrapped on $S^1$.

\bigskip

The system obtained after these dualities will be called the D1D5 system.
As in the case of the NS1-P system, what we want here is the bound state of the D1 branes and the D5 branes. Naively, we might think that when we bind all these branes together then we will get a pointlike mass which we can take to be sitting at the origin $r=0$ of the noncompact space $M_{4,1}$. But our experience with the NS1-P system shows that this might not be right. In the NS1-P case the system had acquired a nontrivial transverse size due to the vibration of the NS1 in the process of carrying the momentum P.  An S duality will not change the transverse size of any system, when we measure this size in the Einstein metric; this is because the Einstein metric is left unchanged by an S duality. T dualities are carried out only in the compact directions, and do not change the transverse size of the system when this size is measured in the {\it string} metric. Thus when we are done with our dualities from NS1-P to D1D5, we will find that the D1-D5 microstate will also have a nontrivial transverse size. The metric (\ref{ttsix}) for the NS1-P system gives, after duality transformations, the following D1-D5 metric \cite{lm4} (the subscript `string' means that the metric is written in the string frame)
\bea
ds^2_{string}&=&\sqrt{H\over 1+K}[-(dt-A_i dx^i)^2+(dy+B_i dx^i)^2]\nn
&+&\sqrt{1+K\over
H}dx_idx_i+\sqrt{H(1+K)}dz_adz_a
\label{qsix}
\eea
where the harmonic functions are
\bea
H^{-1}&=&1+{\mu^2Q_1\over \mu L_T}\int_0^{\mu L_T} {dv\over |\vec x-\mu\vec F(v)|^2}\nn
K&=&{\mu^2Q_1\over
\mu L_T}\int_0^{\mu L_T} {dv (\mu^2\dot
 F(v))^2\over |\vec x-\mu\vec F(v)|^2},\nonumber\\
A_i&=&-{\mu^2Q_1\over \mu L_T}\int_0^{\mu L_T} {dv~\mu\dot F_i(v)\over |\vec x-\mu\vec F(v)|^2}
\label{functionsqq}
\eea
Here $B_i$ is given by
\be
dB=-*_4dA
\label{vone}
\ee
and $*_4$ is the duality operation in the 4-d transverse  space
$x_1\dots
x_4$ using the flat metric $dx_idx_i$.

By contrast the `naive' geometry which one would write for D1-D5 is
\bea
&&ds^2_{naive}={1\over \sqrt{(1+{Q'_1\over r^2})(1+{Q'_5\over
r^2})}}[-dt^2+dy^2]\nn
&&~~\sqrt{(1+{Q'_1\over r^2})(1+{Q'_5\over
r^2})}dx_idx_i+\sqrt{{1+{Q'_1\over r^2}\over 1+{Q'_5\over r^2}}}dz_adz_a\nn
\label{d1d5naive}
\eea

\subsection{The D1D5 CFT}

Suppose that in (\ref{qsix}) we look at a region
\be
\mu |\vec F|\ll r \ll \sqrt{\mu Q_1\over L_T}
\label{slimit}
\ee
Then we see that the metric simplifies to the form
\bea
ds^2_{string}&=&{r^2\over \sqrt{Q'_1Q'_5}}[-dt^2+dy^2]+\sqrt{Q'_1Q'_5}dr^2\nn
&+&d\Omega_3^2 +\sqrt{Q'_1\over Q'_5} dz_adz_a
\eea
This metric has the form
\be
AdS_3\times S^3\times T^4
\ee
Thus we have an asymptotically $AdS$ space if we restrict to the region $r\ll  \sqrt{\mu Q_1\over L_T}$, and we can apply the ideas of AdS/CFT duality. To be able to take the limit (\ref{slimit}) we need that $\mu  \ll \sqrt{\mu Q_1\over L_T}$, and it turns out that this is possible if we take the radius $R$ of the $S^1$ to be large \cite{what}
\be
{R\over \sqrt{Q'_1Q'_5}}\gg 1
\ee
Taking this limit, we expect by Maldacena's AdS/CFT correspondence that there will be a CFT description that is dual to the gravity description. Let us see what this CFT  is.

Recall that we have wrapped $n'_5$ D5 branes on $T^4\times S^1$ 
and $n'_1$ D1 branes on $S^1$. The D1 branes are bound to the D5 branes, so as a first approximation we can say that the D1 branes vibrate inside the plane of the D5 branes. But now note how the corresponding charges behaved in the NS1-P duality frame. Suppose we had a NS1 that was wound $n_1$ times around the $S^1$, which has length $L=2\pi R$. Let us add one unit of momentum P. Then we have
\be
P={2\pi\over L}={2\pi n_1\over n_1 L}={2\pi n_1\over L_T}
\ee
Thus even though we added only one unit of momentum P to the NS1, this unit of momentum looks like $n_1$ units of the basic vibration mode allowed on the full length of the `multiwound NS1'. Let us  call this phenomenon  `{\it fractionation}' \cite{dmfrac}. After duality to the D1D5 frame, we get the following picture. Suppose we have $n'_5$ D5 branes in a bound state. We bind one D1 brane to these D5 branes. Then this D1 brane will appear as a `fractional D1 brane' in the bound state; it will behave as if there were $n'_5$ `fractional D1 branes inside the D5 branes', with the tension of each fractional D1 brane being ${1\over n'_5}$ times the tension of an isolated D1 brane \cite{maldasuss}. If we had $n'_1$ units of D1 charge, then there will be 
\be
N\equiv n'_5n'_1
\ee
 units of `fractional D1 charge' inside the D5 branes. This corresponds to the $n_1n_p$ units of `fractional momentum' that we would find in the NS1-P duality frame.

The D1 and D5 branes each stretch like a `string' along the direction $S^1$.
Now note that  these two kinds of branes can be interchanged by a set of T dualities. Indeed, if we perform  a T duality in each of the four directions of the $T^4$, the D5 branes become D1 branes and the D1 branes become D5 branes. Thus our final model for the D1D5 bound state  should be symmetric under the interchange of these two kinds of branes.   Thus rather than think of having $N=n'_5n'_1$ units of fractional D1 branes inside the D5 branes, we should think of just having $N=n'_1n'_5$ units of an `effective string'  \cite{maldasuss,sw} that winds around the $S^1$. One can advance more rigorous arguments for such a model, but  the above crude picture should suffice for our present discussion.

\subsection{Using the D1D5 CFT}

\begin{figure}[ht]
\includegraphics[scale=.17]{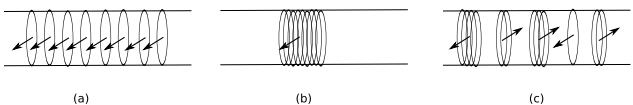}
%
%
\caption{(a) A state with all component strings `singly wound' (b) The state with the entire effective string forming one loop (c) The generic state; there are component strings with many different lengths and spins. }
\label{eff}       
\end{figure}

Let us now see what we can do with this effective string:

\bigskip

(1) {\it Count of ground states:}\quad  First, let us look at all the ground states of this D1D5 CFT. The effective string has total winding number
\be
N\equiv n'_1n'_5
\ee
around the $S^1$. We can have many different configurations with this same total winding. All strands of this effective string could be separate closed loops, as in fig.\ref{eff}(a). Or we could join them all into one long string, as in fig.\ref{eff}(b). More generally,  we would get $m_k$ strands with winding number $k$, with 
\be
\sum_k k m_k =N
\label{ssum}
\ee
Counting all these different possibilities gives $e^S$ states, with
\be
S=2\sqrt{2}\pi\sqrt{N}
\ee
This agrees with (\ref{entropy}), as it should, since the D1D5 system is the same as the NS1-P system under S,T dualities. 

\bigskip

(2) {\it Identifying CFT states with gravity solutions:}\quad We have made D1D5 gravity solutions in (\ref{qsix}), and sketched the CFT states in fig.\ref{eff}. But which CFT state corresponds to which gravity solution?

The link is made by going through the solutions in the NS1-P language. Start with the D1D5 CFT state, and look at a loop with winding number $k$. Each separate such loop is called a `component string'. In the NS1-P picture the string state (\ref{sstate}) was described by oscillators acting on the vacuum state. The oscillator  $\hat a^{i\dagger}_{k}$  maps to a component string with winding number $k$. The polarization $i$ of the vibration gives a `spin' for the component string, which we have drawn with arrows in fig.\ref{eff}. 

Thus start with a CFT state, find the corresponding NS1-P state from (\ref{sstate}) and  find the profile function $\vec F$ for these vibrations of the string. Putting this $\vec F$ in (\ref{ttsix}) gives the metric of this NS1-P state, and performing S,T dualities gives the metric (\ref{qsix}) in the D1D5 duality frame. This then is the metric dual to the CFT state that we started with. As mentioned above, to get a well defined profile function $\vec F$ we need large occupation numbers $m_k$ for each $k$ in (\ref{ssum}). If this is not the case, we get quantum fluctuations and the system is not well described by a classical geometry; this gives the general `fuzzball' configuration.  (The reader can consult the references \cite{review1} for details on the approximations needed to get a classical geometry and for more details of the CFT-gravity map  \cite{skenderis}.) The essential point property of fuzzballs is their size and not how `quantum' the solution is.   As noted in (\ref{ssize}) the {\it size} of the generic state is order horizon size; how `quantum' this state is depends on whether the excitations of the NS1 are concentrated into a few harmonics or spread over many harmonics.

\bigskip

(3) {\it Energy gaps:}\quad So far we have looked at ground states of the D1D5 CFT. Let us now add some extra energy to one of these ground states, making a non-extremal state. The dynamics of the effective string is very simple if we are at `weak coupling': we just get massless bosonic and fermionic modes travelling up and down the effective string (these are called left and right moving modes respectively). While the CFT should actually be at strong coupling to reflect the gravity solution, we will use it at weak coupling where we can actually compute, and hope that the corrections are not large since we are `close' to supersymmetric configurations.

\begin{figure}[ht]
\includegraphics[scale=.18]{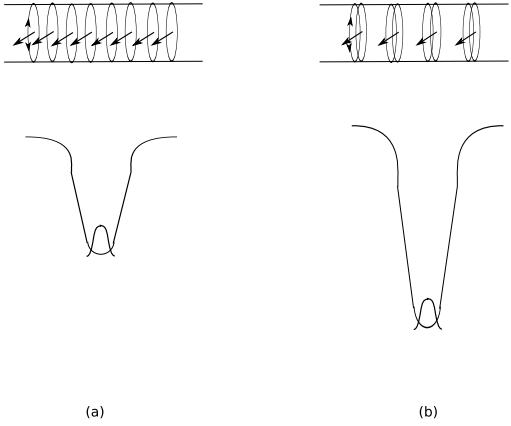}
%
%
\caption{(a) The lowest energy excitation for the CFT state, and its gravity dual; the energies agree in the two descriptions (b) The same computation for a different microstate; the energies again agree between the two descriptions, but are half of the energies in (a). }
\label{effex}       
\end{figure}

In fig.\ref{effex}(a)  we take the state where all component strings are singly wound, and add an excitation on one component string; let this excitation be in the lowest harmonic allowed on the component string. This is the lowest energy excitation of this CFT state, and has an energy $(\Delta E)_{CFT}$. In the gravity dual, we see that we can place a quantum in a wavefunction at the bottom of the throat. Let the lowest allowed energy for such a quantum be  $(\Delta E)_{gravity}$. One finds
\be
(\Delta E)_{CFT}=(\Delta E)_{gravity}
\label{sagree}
\ee
In fig.\ref{effex}(b) we take the CFT state with winding number $k=2$ for each component string. The lowest allowed excitation energy is now {\it half} the value in fig.\ref{effex}(a). But the corresponding gravity dual has a deeper throat; this makes the quantum in the geometry suffer a larger redshift, and we again get (\ref{sagree}).

We see from this analysis that we {\it must} have `caps' for the geometries dual to the D1D5 CFT states. If we had the naive geometry of fig.\ref{geometries} (a) or (b), then we would not get agreement of energy gaps between the CFT and gravity  pictures. 

\bigskip

\begin{figure}[ht]
\includegraphics[scale=.16]{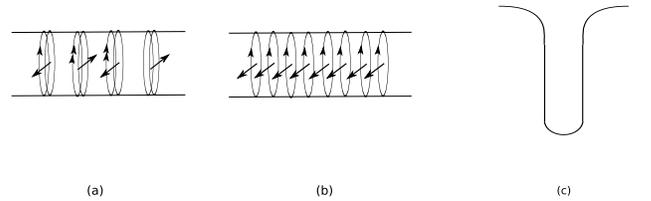}
%
%
\caption{(a) The generic 3-charge extremal CFT state (b) A simple CFT state (c) The geometry for the state in (b) can be explicitly constructed; it has no horizon, and ends in a smooth `cap'. }
\label{threec}       
\end{figure}

(4) {\it 3-charge geometries:}\quad If we add excitations carrying momentum P {\it up} the component strings, but not {\it down}, then the state get a net momentum charge P which equals the energy added. We then get states of the 3-charge extremal hole \cite{sv}. The generic CFT state of this hole is pictured in fig.\ref{threec}(a). We do not yet know how to make the dual of a generic 3-charge CFT state. But let us look at the simple 3-charge state depicted in fig.\ref{threec}(b); because all the component strings have equal length and spins, the geometry has axial symmetry, and we {\it are} able to construct the gravity dual. This dual is given by the metric \cite{gms1,gms2,lunin}
\bea
&&ds^2 = -\frac{1}{h} (dt^2-dy^2) + \frac{Q_{p}}{h f}\left(dt-dy\right)^{2} \nn
&&~~~~+ hf
\left( \frac{dr^2_N}{r_{N}^2 + a^2 \eta } + d\theta^2 \right)\nonumber \\
&& +h \left( r_{N}^2 - n a^{2}\eta + \frac{(2n+1)a^2\eta Q_{1}Q_{5} \cos^2\theta}{h^2 f^2} \right)\hspace{-.04 true in}
\cos^2\theta d\psi^2   \nonumber \\
&&+ h\left( r_{N}^2 + (n+1) a^2 \eta - \frac{ (2n+1)a^2\eta Q_{1}Q_{5} \sin^2\theta}{h^{2} f^{2} }\right)\nn
&&~~~~~~~~~~~~~~~~~~~~~~~~~~~~~~~~~~~~~~~~~~~~~~~~~~~~~~~~~
\sin^2\theta d\phi^2  \nonumber \\
&&~+ \frac{a^{2} \eta^2 Q_p}{h f} \left( \cos^2\theta d\psi + \sin^2\theta d\phi \right)^{2}  \nonumber\\
&&~+ \frac{2 a \sqrt{Q_{1}Q_{5}} }{hf} \left[ n \cos^2\theta d\psi - (n+1) \sin^2\theta d\phi\right]\hspace {-.05 true in} (dt-dy) 
\nonumber \\
&&~- \frac{2 a \eta \sqrt{Q_{1}Q_{5}}}{h f} \left[ \cos^2\theta d\psi + \sin^2\theta d\phi \right] dy \nn
&&~+
\sqrt{\frac{H_{1}}{H_{5}}} \sum_{i=1}^{4} dz_{i}^2 \nn
\label{em}
\eea
where
\bea
\eta&\equiv& {Q_1 Q_5\over Q_1 Q_5 + Q_1 Q_p + Q_5 Q_p}\nn
f &=& r_{N}^2 - a^2\eta \,n \sin^2\theta + a^2 \eta\, (n+1) \cos^2\theta \nonumber\\
h &=& \sqrt{H_{1} H_{5}}, \ H_{1} = 1+ \frac{Q_{1}}{f}, \ H_{5} = 1+ \frac{Q_{5}}{f}
\label{deffh}
\eea
Again one finds that there is no horizon, and the geometry ends in a smooth `cap' (fig.\ref{threec}(c)). The energy gaps for the 3-charge CFT states agree exactly with the energy of quanta placed in the geometry fig.\ref{threec}(c). These facts suggest very strongly that all we have learnt for 2-charge extremal holes (where we can understand  all states) will also hold for 3-charge extremal holes.

\bigskip

\begin{figure}[ht]
\includegraphics[scale=.11]{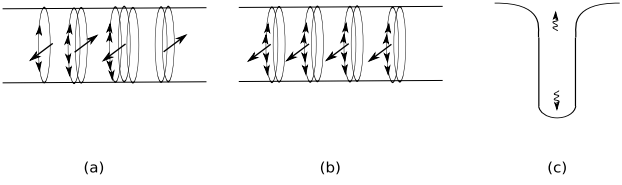}
%
%
\caption{ (a) The generic nonextremal state in the CFT (b) The special states that we consider (c) The geometry for the special states; pair creation occurs near the ergoregion, one member of the pair falls into the ergoregion, while the other escapes to infinity.  }
\label{nonex}       
\end{figure}

(5) {\it Non-extremal holes:} \quad We have seen in (2) above that we get non-extremal states if we have excitations running both up and down the string. In the case (2) we added only one excitation to one component string, so in the gravity dual we had just one quantum sitting in the geometry. We could ignore the backreaction of this single quantum, and so solved the free wave-equation on the extremal background. Let us now consider the general non-extremal state, where we have an arbitrary number of left and right excitations on the component strings. We depict the general state of the non-extremal system in fig.\ref{nonex}(a). If we could understand the gravity dual of such CFT states, we would have understood the non-extremal black hole. 

We cannot construct the gravity duals of the generic states  fig.\ref{nonex}(a), but we {\it do} know how to make duals of special states like the one in fig.\ref{nonex}(b). All the component strings have been chosen to have the same length and spins; further, the left and right excitations are all fermionic and chosen so that they occupy the lowest allowed levels for these fermions. In this case the gravity dual is found to have the structure \cite{ross} 
\begin{eqnarray} \label{2charge}
&&ds^2=-\frac{f-M}{\sqrt{\tilde{H}_{1} \tilde{H}_{5}}}
dt^2+\frac{f}{\sqrt{\tilde{H}_{1} \tilde{H}_{5}}}dy^2\nn
&&+\sqrt{\tilde{H}_{1} \tilde{H}_{5}}
\left(\frac{ dr^2}{ r^2+a_{1}^2 - M}
+d\theta^2 \right)\nonumber \\ 
&&+\left( \sqrt{\tilde{H}_{1}
\tilde{H}_{5}} + a_1^2 \frac{( \tilde{H}_{1} + \tilde{H}_{5}
-f+M) \cos^2\theta}{\sqrt{\tilde{H}_{1} \tilde{H}_{5}}}  \right)\nn
&&~~~~~~~~~~~~~~~~~~~~~~~~~~~~~~~~~~~~~~~~~~~~~~~~ \cos^2
\theta d \psi^2 \nonumber \\ 
&& +\left( \sqrt{\tilde{H}_{1}
\tilde{H}_{5}} -a_1^2 \frac{(\tilde{H}_{1} + \tilde{H}_{5}
-f) \sin^2\theta}{\sqrt{\tilde{H}_{1} \tilde{H}_{5}}}\right) \sin^2
\theta d \phi^2 \nonumber \\ 
&&+ \frac{2M \cos^2 \theta}{\sqrt{\tilde{H}_{1} \tilde{H}_{5}}}(a_1
c_1 c_5 ) dt  d\psi +\frac{2M \sin^2 \theta}{\sqrt{\tilde{H}_{1} \tilde{H}_{5}}}(a_1
s_1 s_5 ) dyd\phi \nn
&&+ \sqrt{\frac{\tilde{H}_1}{\tilde{H}_5}}\sum_{i=1}^4
dz_i^2 \label{Eqn:metric}
\end{eqnarray}
where
\be
c_i = \cosh \delta_i, \quad s_i=\sinh \delta_i
\ee
\begin{eqnarray} 
\tilde{H}_{i}=f+M\sinh^2\delta_i, \quad
f=r^2+a_1^2\sin^2\theta, \label{Def:FandH}
\end{eqnarray}
The geometry again has no horizon, and is sketched schematically in fig.\ref{nonex}(c). 

It is exciting that we have been able to make non-extremal states and found them to also be `fuzzballs' rather than `metrics with horizon'. But more is true. We can also study Hawking radiation from these non-extremal states.

First consider the generic CFT state in fig.\ref{nonex}(a). The left and right moving excitations can collide and leave the CFT bound state as radiation. The rate of this process is given by an emission vertex $V$ times the occupation probabilities for the left and right colliding modes.  Symbolically, 
\be
\Gamma = V\rho_L\rho_R
\ee
If we put thermal distributions for $\rho_L,\rho_R$, then $\Gamma$ agrees exactly with the Hawking emission from the near-extremal black hole \cite{radiation}
\be
\Gamma=\Gamma_{Hawking}
\ee
Of course here we have agreement only of the radiation {\it rate}, not the details of emission. The CFT emission $\Gamma$ is a unitary process in a normal thermodynamic system, while $\Gamma_{Hawking}$ is the semiclassical computation in the black hole geometry which leads to information loss.

Let us see if we can do better with our understanding of fuzzballs. We cannot yet make the gravity dual of the general state fig.\ref{nonex}(a), but let us see if we can understand emission from the special states fig.\ref{nonex}(b) that we {\it can} make. In the CFT description we get the emission by replacing the occupation numbers $\rho_L, \rho_R$ with the ones appropriate to this special microstate 
\be
\Gamma_{CFT}=V\bar\rho_L\bar\rho_R
\ee
On the gravity side, we find that the geometry (\ref{2charge}) is {\it unstable}, and radiates energy out to infinity \cite{myers}. The rate of this radiation is found to exactly agree with the rate of emission from the CFT \cite{cm1}
\be
\Gamma_{gravity}=\Gamma_{CFT}
\ee
With such an explicit description of the emission from the gravity state, we can ask how and where the radiation arises. The geometry of the microstate has no horizon, but it does have an ergoregion. Thus we get the process of ergoregion emission, whereby particle pairs are produced near the ergoregion; one member of the pair falls into the ergoregion while the other escapes to infinity as radiation. But the member that falls in is not `lost' as would be the case for traditional Hawking radiation; instead it influences the production of further quanta from the ergoregion. This happens because of a `Bose enhancement' process; after $n$ quanta have collected in the ergoregion the probability to create the next quantum is proportional to $n+1$. The emission thus increases exponentially, and is  characterized by a set of complex frequencies
\be
\omega^{(i)gravity}=\omega^{(i)gravity}_R+i\omega^{(i)gravity}_I
\ee
In the dual CFT state fig.\ref{nonex}(b) we also find emission peaked at certain discrete frequencies since we have taken all component strings to be excited in  the same way. We again find an exponential growth of emission, with complex frequencies in exact agreement with the gravity emission \cite{cm1}
\be
\omega^{(i)CFT}=\omega^{(i)gravity}
\ee
The emission from our special microstates is peaked at special frequencies like a laser instead of being the planckian emission spectrum expected from warm bodies. But this is of course expected; each microstate emits somewhat differently, and if we start with a very special microstate where all excitations are at a given energy then we will get a peculiar emission behavior. The important point is that we get exact agreement between the  CFT computation  and a gravity calculation which this time {\it gives the same emission by a unitary process with no information loss}. In particular, we see that the quanta that fall into the ergoregion  influence the production of the next quantum through Bose enhancement. This should be compared to the discussion of section \ref{introduction} where we noted that radiation from a piece of coal can carry out information because radiated quanta can `see' the effects of earlier radiated quanta, while in the traditional computation of Hawking radiation the newly produced pairs do not see the state of earlier produced pairs. 

\section{Towards making all extremal fuzzball states}

The 2-charge extremal hole requires $R^2$ corrections at its horizon to get the exact Bekenstein-Wald entropy. Thus while we can understand all states of the 2-charge hole, we would  like to study 3-charge and 4-charge  extremal holes, which have a larger horizon and do not require such corrections.

\subsection{3-charge and 4-charge states}

For 3-charge and 4-charge extremal  holes we do not yet have a systematic way of constructing all states in the gravity description. But for all those states which have been constructed, we find that we get `fuzzballs': the throats are finite and capped, not infinite and ending in a horizon.

The simplest 3-charge extremal states are those with $U(1)\times U(1)$ axial symmetry; these states were constructed some years ago  \cite{gms1, gms2, lunin}.
How do we make more general 3-charge solutions? It can be shown that any  supersymmetric solution for N=1 supergravity in 6-d can be written as a 2-d fiber over a hyperkahler base \cite{gmr}.
The $U(1)\times U(1)$ extremal solutions \cite{gms1,gms2} can be dimensionally reduced on the $T^4$ to give solutions in 6-d, and we can then ask what this base-fiber split looks like. Interestingly, the base turns out to be `pseudo-hyperkahler': the signature of the base jumps from being $(++++)$ to $(----)$ across a hypersurface in the base \cite{gm1}. The fiber degenerates at this hypersurface too, in such a way so that the overall 6-d metric remains smooth. Thus the lesson is that while local supergravity equations tell us that the solution will have a hyperkahler base and a 2-d fiber, in the actual solutions corresponding to D1-D5-P extremal states this split cannot be performed globally; it degenerates along certain surfaces.

In a very interesting series of papers \cite{bena}, Bena and Warner took this story to a new level. They started from the equations of 11-d M-theory, and obtained a more detailed version of this base-fiber split. Specializing the hyperkahler base to Gibbons-Hawking spaces (which have an extra U(1) symmetry), they managed to get a complete solution of the supergravity field equations. The fact that the space was pseudo-hyperkahler (rather than hyperkahler) could be easily built into their formalism: the solutions were written in terms of harmonic functions on the base, and the sign of the sources in these harmonic functions  determined the local signature of the base. With this formalism, it became possible to write down explicitly large families of supersymmetric solutions to string theory, all having the mass and charges of the 3-charge black hole. None of the solutions had a horizon or `black hole singularity'. The sources of the harmonic functions are held apart at fixed distances by fluxed running on spheres joining them; these constraints are given by `bubble equations', which contain the essence of the supergravity equations in the present ansatz.

The above mentioned solutions had one $U(1)$ symmetry -- the one needed to make the pseudo-hyperkahler base a Gibbons-Hawking space. The solutions have  4+1 noncompact dimensions. We can do a dimensional reduction along the circle corresponding to the remaining $U(1)$ symmetry, thus getting solutions in 3+1 noncompact dimensions. The way to do this compactification  is to make the circle  the fiber of a Kaluza-Klein monopole. The solutions acquire a fourth charge, that of the KK monopole, and we get 4-charge solutions in 3+1 noncompact dimensions. (Note that if we want to make an extremal black hole with classical horizon size in 3+1 dimensions, then we have to use four charges.) Such solutions have recently been  constructed  \cite{giusto,gimon,bena}. The bubble equation in this 3+1 dimensional setting become similar to equations studied earlier by Denef \cite{denef}. In fact Denef had developed an elegant  general formalism for making supersymmetric solutions out of more fundamental constituents. These fundamental constituents could be individual branes (having no entropy) or extremal black holes (having a nonzero entropy). The fuzzball proposal would say that all states of the system can be written in terms of constituents {\it without} entropy. It is not clear if the elementary constituents used in the references above \cite{gimon} are `complete'; it is likely that there are more complicated constructions that need to be taken into account before we have all 4-charge extremal states.

A general philosophy that emerges from all these constructions is the importance of `dipole charges'. The supersymmetric solutions have some charges that we measure from infinity; let us call these the `true charges' of the solutions. When we look at the actual microstate solutions, we find that we have flat space at infinity, then for some distance we have  the uniform throat expected of the traditional black hole, and then a `cap' region. In this cap we find, besides the `true' charges,  a set of charges that are not measured as charges at infinity. These are `dipole charges' and their net value adds up to zero. But their locations can be varied, and this gives us different solutions corresponding to the same total mass and `true' charges. Exploring the space of such allowed solutions is therefore relevant to exploring the structure of general black hole microstates. (One can think of these dipole systems as supertubes with more than two charges; for some generalization of supertubes to three or more charges, see \cite{super2,super3}.)

With this wealth of available tools, a large variety of supersymmetric solutions have been made for the 3-charge and 4-charge cases. One can make structures that look like microstates of holes, or rings, or a collection of holes and rings. Some choices of fluxes lead to `deep throat' solutions, which may account for a large fraction of the microstates of the hole. Solutions depending on a continuous parameter were recently found \cite{bwrecent} by putting a supertube inside a deep throat. With such a construction it may be possible to get enough solutions that their number will go like $ Exp[\sqrt{n_1n_2n_3}]$ for charges $n_1, n_2, n_3$; in that case one would have an entropy from these solutions that would account for the black hole entropy, and we would be in a situation similar to the one that we had for the 2-charge case.

Several other studies have been done with   extremal solutions. Steps have been taken to quantize the moduli space of these solutions \cite{deboer},  to study the mathematical properties of the family of such solutions  \cite{verlinde}, and to coarse grain over the solutions to get an `entropy' \cite{bhentropy}. 

\subsection{`Hybrid' models}

While we could make the gravity states of the 2-charge extremal system with comparative ease, we have seen that it is  hard to make the gravity duals for general states of the three and four charge extremal holes. One approach in this situation has been to treat some of the charges `exactly', finding their exact gravity description, while letting the other charges be placed as a small perturbation in the background produced by the first set of charges. With such an approach we may be better able to think of the complete ensemble of all states, though we will lose some understanding of the full gravity description of the state since some of the charges have not been handled with full backreaction. Let us see how some of these approaches proceed.

Since the entropy of a black hole is given by its surface area, it has always been tempting to find some degrees of freedom that live at the horizon and whose count gives the entropy of the hole. The problem with this of course is that we cannot place something at the horizon and expect it to stay there; any excitation at the horizon either falls into the hole or escapes to infinity, leaving no degrees of freedom at the horizon. This is just the standard `no hair' phenomenon found for traditional black hole geometries, and has been a long standing problem in understanding the entropy of black holes. 

But now we have learnt that at least for simple cases of extremal black hole states, we do not have a horizon but instead a geometry that `caps off' before a horizon is reached. In the simplest case of the 2-charge extremal D1D5 solution, the profile function $\vec F$ is the helix sketched in fig.\ref{strings}(a). In this case the cap region has the geometry of global $AdS_3\times S^3\times T^4$.
Thus let two of the charges making the hole be D1 and D5, and let these charges be in a state which generates this particular 2-charge geometry. Now let us add other excitations as perturbations, creating new excitations that we can count but for which we will not take the gravitational backreaction into account. What excitations should we take? It has been noted \cite{mandal} that we can put `giant gravitons' \cite{giant} in $AdS$ type geometries. These giant gravitons are branes which wrap spheres in the $AdS$ space or the sphere, and are preventing from collapsing to a point because they move through the gauge field flux which exists in the background geometry. Counting these giant gravitons one finds enough states to account for the entropy of a 3-charge hole. Note however that  since we have not considered the gravitational backreaction of these giant gravitons we cannot  say that we understand the full gravity description in this picture.

A counting has been suggested \cite{strominger} for `brane states wrapping  a black hole horizon'.  The count gives  a number that agrees with the entropy of the corresponding hole.  For the reasons mentioned above, it is completely clear where and how such brane states would be located in the presence of the  horizon. TRo see if this count could be put on a firmer footing,   an attempt was made \cite{ok} to understand such a counting of branes by replacing the effect of some of the charges by the capped geometry they would produce; the other branes were then put as test charges in the capped geometry. There is no horizon now, but the branes wrapped a sphere which is analogous to the spherical horizon in the naive black hole geometry. With this construction the  branes did not fall through a horizon, and thus could be localized and counted. But a different problem emerged. The branes wrapping the sphere turn out to act like `domain walls', so that the value of the flux they produced jumps from one side of the wrapped brane to the other. Regularity in the cap required no field on the `inner' side of the brane, so one gets a nonzero field on the `outer' side which extends all the way to infinity. Thus wrapping a brane in this fashion on a sphere produces a nonzero  gauge field strength over an infinite volume, making the state have infinite energy. Perhaps some other method of wrapping branes may be more appropriate to counting the degrees of freedom of the system.

The 4-charge hole has been studied similarly \cite{denefstrom} by letting some charges form a `capped' background, and letting the other charges be added as test branes. Interactions between these test charges were also considered, and it seemed possible that D0 branes placed in the background geometry could swell up to D2 branes wrapping spheres by the Myers mechanism \cite{myersp}. The construction has been  extended to black rings  \cite{gimonrings}. It should be checked though if in all these cases one can avoid the above mentioned problem of infinite flux energies.

\section{Dynamical processes: conjectures}\label{dynamical}

The fuzzball program constructs the states of the black hole, in the gravity description. These states can be thought of as energy eigenstates of the system. Thus they  do not individually describe time dependent processes like the formation of a black hole by collapse of a shell. But once we understand the energy eigenstates of a system, we can reconstruct its dynamics by superpositions of these eigenstates. While we do not have a comprehensive picture of all non-extremal microstates, in this section we will try to conjecture some aspects of the dynamics that should result if all black hole microstates were indeed fuzzballs. 

The main dynamical questions of interest are of the following type. What happens to an observer as he approaches the horizon?  How should we understand his evolution inside the hole? How does his information come out in the Hawking radiation? If we start with a collapsing shell, how does it evolve into a fuzzball? Let us consider what we have learnt about black hole microstates and see if we can postulate how some of these questions might be answered.

\subsection{The two scales in black hole physics}

A common first question about fuzzballs is the following. In the traditional picture of the hole we have vacuum at the horizon, so an infalling observer feels nothing as he crosses the horizon. In the fuzzball the information of the hole is distributed throughout a horizon sized ball. So will the observer feel something drastically different as he approaches the place where he expected a horizon?

To understand this and similar issues, 
it is important to note that there are two different time scales of interest in the black hole problem. One is the `crossing time scale' $t_{cross}$ over which an infalling quantum travels from the horizon to the singularity. The other is the much longer Hawking evaporation timescale $t_{evap}$, which for a 3+1 dimensional Schwarzschild hole is $({M\over m_{pl}})^{2}$ times $t_{cross}$. Thus we can say that $t_{evap}$ is larger than $t_{cross}$ by a power of ${1\over \hbar}$. 

Now consider a quantum falling into the hole. The density of the `fuzz' for a generic state of the hole was computed \cite{phase}, and found to be low at the horizon. Thus there need not be a sharp interaction of the infalling quantum with the degrees of freedom of the hole; in fact there is no contradiction in assuming that the motion of the quantum over the time $t_{cross}$ resembles the free fall in the traditional black hole geometry. What we {\it need}  to solve the information problem is that the interaction of the infalling quantum with the degrees of freedom of the hole happen in a time smaller than $t_{evap}$, so that the information of the quantum can indeed come out in the Hawking radiation. Since $t_{evap}\gg t_{cross}$, there is no contradiction in assuming very different evolutions on these two different time scales.

The existence of these two different scales makes it possible to preserve some part of our classical intuition about black holes while resolving the information puzzle. This could be part of a more general principle. In the extremal hole, we see two different {\it length} scales. In the traditional extremal geometry the throat has an infinite length. In the fuzzball picture, the length of the throat for a generic 3-charge geometry has been estimated  \cite{higher}. Suppose the diameter of the throat is $D$. the length of the throat is then a power of the charges $n_{1}n_{5}n_{p}$ times $D$. From a macroscopic perspective, we can say that the depth of the throat is a power of ${1\over \hbar}$ times its diameter.  Thus if we look only a down the throat only upto a fixed multiple of $D$   then for $n_i\rightarrow\infty$  we will see just the classical throat geometry and not the quantum fuzz at the end of the throat.

These computations suggest a `classical correspondence principle', which would say that to leading classical order the fuzzball states behave in a way expected from the traditional hole. We do not yet have a clear formulation of such a principle, but let us note some other computations which might help formulate such a principle.

Consider the 2-charge extremal geometries. If we take a simple geometry like the one pictured in fig.\ref{strings}(a), then an infalling quantum bounces off the end and returns back in a small time. Now consider the geometry for a generic state (fig.\ref{geometries}(c)). This geometry is very complicated in the `cap region, and an infalling quantum will be trapped in that region for a long time.  was estimated The time of return from a generic 2-charge geometry was found \cite{lm5} to be a power of $n'_{1}n'_{5}$ times the crossing time across the fuzzball; this long time results from the many deflections a geodesic suffers before it can exit the cap region. Again, we can think of this return time as a power of ${1\over \hbar}$ times the crossing time. For an observer who looks at the system only for a fixed multiple $\mu$ times the crossing time, the infalling quantum would appear to be lost for ever when we take the charges to infinity. Thus for such an observer we can  replace the boundary of the fuzzball by a traditional horizon, and obtain essentially the same effect: now the infalling quantum would never return. It is in this sense that we should understand the emergence of a horizon in the fuzzball picture. The `horizon' is only an effective concept describing the evolution over the short timescale $t_{cross}$, while the actual details of the quantum fuzz lead to the eventual leakage of information from the fuzzball, something that cannot happen if we {\it really} had the traditional black hole horizon. 

Just as we differentiated between two different time scales and length scales, we should also separate two different {\it energy} scales. The typical Hawking radiation quantum has an energy $E_{Hawking}$ of order the temperature $T$ of the hole. When we think of an infalling observer, we should ask if the energy of this observer is $E\sim E_{Hawking}$, or if  $E\gg E_{Hawking}$. From our analysis of the information paradox we know that the evolution of Hawking radiation quanta with $E\sim E_{Hawking}$ {\it must} be modified by order unity by the detailed information in the fuzzball state; otherwise information will not come out in the radiation. On the other hand, it is not necessary that the evolution of modes with $E\gg E_{Hawking}$ be affected to leading order by the fuzzball structure, at least for time scales $t\ll t_{evap}$.

As an explicit example of this, consider the nonextremal geometry that we considered in the last section. The Hawking emission happens because of the negative effective potential in the ergoregion, and this emission does not happen from the part of the geometry which is not in the ergoregion. But the negative potential is quite small, and the emitted quanta  have a low energy. If we send a high energy quantum into the geometry, it does not notice the ergoregion potential in any significant way, and its evolution does not depend sensitively on whether or not it passes through the ergoregion. Thus here we have a simple example where the evolution of the Hawking radiation quanta depends on sensitive details of the geometry while the evolution of a `heavy' infalling observer is not sensitive to the same details.

\subsection{Formation of fuzzballs}

If the energy eigenstates of the black hole are horizon sized fuzzballs, then any infalling shell should eventually be best described by  a linear combination of these fuzzball geometries. But how will this happen? A classical shell seems to feel no large quantum effects as it crosses the horizon, so one would think that the result should be the traditional black hole with the `information free horizon'. In this section we will make some simple observations which indicate why black holes may not be as classical as they at first appear.

\subsubsection{Tunneling between macroscopic states}

Consider any state of matter which has mass  $M$ , and which is localized in a region $R\sim GM$ which is order the  black hole radius for mass $M$. A collapsing shell would be such a state as it crosses  its horizon. Now consider any other state which has the same mass and which is localized in the same region, for example a fuzzball state.
 Let us ask if there is any significant amplitude for `tunneling' between such states, postponing for the moment the details of what this tunneling process is. (We will see below that we are looking for `spreading of a wavefunction' rather than tunneling, but it is more helpful to think of a tunneling process on a first pass at the issues.)  

Normally the tunneling amplitude would be small, since the states have large mass and size. We will estimate the action for a tunneling process by writing
\be
S_{tunnel}\sim {1\over G} \int R d^{4}x\sim {1\over G} {1\over (GM)^{2}}(GM)^{4}\sim {GM^{2}}
\label{saction}
\ee
where we have assumed a length scale $GM$ for the curvature and a volume $(GM)^{4}$ over which the process takes place. Thus the amplitude for tunneling from the shell to a fuzzball state  
\be
{\cal A}\sim e^{{-S_{tunnel}}}
\ee
is very small.

But now note that there are a very large number of fuzzball states that 
we can tunnel to. This number is given by
\be
{\cal N}\sim e^{S_{bek}}\sim e^{GM^{2}}
\ee
We see that something curious happens for black holes. These objects have such a large entropy that the very small probability for tunneling between classical configurations can be compensated for by the very large number of states that we can tunnel to \cite{tunnel}. This would make a black hole an essentially quantum object. Note that if we took a star instead, then the action (\ref{saction}) is larger (the size of the object is bigger) while the entropy is much lower, and there is no such quantum behavior.

\subsubsection{Spreading over phase space}\label{spread}

The above was just a crude order of magnitude estimate, but now let us see if we can say something more about the actual dynamical process of shell collapse. The crucial point will be the fact that there is a large number of possible states states of the hole -- the $e^{S_{bek}}
$ fuzzballs. In the classical picture of collapse we do not see these  states which are supposed to give the entropy of the hole. We will see that it may not be correct to ignore the large phase space which these microstates represent, and when we do take all these solutions into account the quantum evolution of a collapsing shell can be very different from its classical approximation.

Let us proceed in three steps.

\bigskip

(1) First let us take a 2-charge extremal geometry, and throw into the throat a quantum of a scalar field $\phi$  with energy $E$. We choose $E$ to be small, so the backreaction of the quantum on the geometry can be ignored. The quantum will fall down the throat,  reach the cap, and eventually reflect back up the throat. How do we describe this evolution in terms of the energy eigenstates of the system?

We can find the energy eigenstates of the quantum by solving the wave-equation $\square \phi=0$. (For the simple geometries of fig.\ref{effex} the wavefunctions have been explicitly computed \cite{ppwave}.) We get a set of energy eigenfunctions. The lowest energy state is localized in the cap (as shown in fig.\ref{effex}), the next one extends a little further out, the next one still further, etc. The infalling quantum starts high up the throat, so we must superpose these energy eigenfunctions with suitable coefficients to obtain this initial wavepacket
\be
|\psi\rangle=\sum_k c_k |E_k\rangle
\label{sum}
\ee
where $|E_k\rangle$ is the eigenfunction with energy $E_k$. 

This is all just standard quantum mechanics, and we would do a similar  computation for describing a  localized quantum moving in the potential of a harmonic oscillator. The evolution of the wavepacket down the throat is obtained by evolving the energy eigenfunctions; since these eigenfunctions have slightly different energies, the relative phases between their coefficients change with time and cause the wavepacket to move downwards towards the cap. The essential point in the above discussion is that even though the quantum is localized quite high up the throat up the start, if we want to express its wavefunction in terms of the stationary states of the system then we have to construct the detailed energy eigenfunctions $|E_k\rangle$ in the entire geometry, and these will depend sensitively on the structure of the cap.

\bigskip

(2) Now let us imagine that the energy of the infalling quantum is a bit higher. We would therefore like to take into account the small backreaction that the infalling quantum would create on the geometry. How should we do this? 

We still have to follow the same basic scheme: we have to find the energy eigenstates of the system and  superpose them with appropriate coefficients.  The evolution will then be given by the changing phases of the coefficients. But what are the energy eigenstates this time? Clearly, we should find solutions to the full system of gravity plus scalar field $\phi$, with the backreaction of the $\phi$ excitation included, and arrive at some eigenstates $\psi_k[g, \phi]$ which are functionals of both the metric $g$ and the scalar field $\phi$. Note in particular that the energy $E_k$ of this state will reflect the energy of the background extremal 2-charge geometry as well as the energy of the quantum.  So we are making energy eigenstates around an energy 
\be
E_{total}=E_{extremal}+E_{quantum}
\label{stotal}
\ee
The number of states of the system increase with the energy, and we observe here that the set of eigenstates that will be involved in a sum like (\ref{sum}) will be the number at energy $E_{total}$, and not at the base energy $E_{extremal}$.

\bigskip

(3)  Now let us imagine increasing the energy of the infalling quantum still further, so that a classical analysis would indicate the formation of a horizon at some point in the throat, much before the cap is reached.
This is of course the case that we are really interested in understanding. The basic scheme will remain the same as in the above two cases, but now we have to find all energy eigenstates of the system with an energy $E_{total}$  where the contribution $E_{quantum}$ is not small. According to our postulate, these energy eigenstates are horizon sized fuzzballs, pictured in fig.\ref{geometries}(c). Thus the initial infalling quantum has to be written in the form (\ref{sum}) as a set of very quantum fuzzball states; these states are very numerous and have a nontrivial structure all the way upto the horizon. 

Now suppose we did not know that there were all these fuzzball states, and we wrote the sum (\ref{sum}) with only the states that we see in the traditional picture of the black hole. Then we would be using a  much smaller number of states. For example if we took the infalling quantum to have spherical symmetry, then we might (erroneously) assume that the black hole background should be a classical spherically symmetric state. But from what we have seen of fuzzball states, they are in general {\it not} spherically symmetric. Spherical symmetry of the overall state is obtained by superposing with equal coefficient a {\it non-spherical} geometry with all of its rotates. Thus if we write the initial shell as a superposition of spherically symmetric fuzzball states, then these states will have large {\it fluctuations} ${\delta g\over g}$.

In short, the fuzzball picture would give a much larger sum of states in (\ref{sum}) as compared to a traditional picture which does not explicitly recognize the degrees of freedom corresponding to the Bekenstein entropy. As the phases of the coefficients $c_k$ evolve, the initial state with the quantum will change to a general linear superposition of fuzzball states, something we cannot see in the traditional classical infall.

It is interesting to note the phase evolution of the $c_k$ becomes important in a time that is shorter than the Hawking evaporation time. Suppose we have a shell of mass $M$ that collapses to form a black hole. Let the Schwarzschild radius of the hole be denoted by $R$. To make the shell collapse we must localize the matter in the shell so that it fits in a radius $\ll R$. This needs a momentum spread for the shell
\be
\Delta P\gg  {1\over R}
\ee
For a nonrelativistic shell, the energy of the shell is $E\sim {P^2\over 2M}$, and the uncertainty in $E$ will; be
\be
\Delta E\sim {P\Delta P\over M}\gg {(\Delta P)^2\over M}\gg {1\over MR^2}
\ee
The different fuzzball states $|E_k\rangle$ making up the shell wavefunction $|\psi\rangle$ will go `out of phase' over a time $t_{dephase}$ so that the state will look like a linear combination of generic fuzzball states rather than a well defined shell. We have
\be
t_{dephase}\sim {1\over \Delta E}\ll MR^2
\ee
But the Hawking evaporation time for a Schwarzschild hole (in all dimensions) is
\be
t_{evap}\sim { MR^2}
\ee
Thus we find that the time over which the the wavefunction `dephases to fuzzballs' is shorter than the Hawking evaporation time
\be
t_{dephase} \ll t_{evap}
\ee
 This is important, since this `dephasing' would not be of interest if it took {\it longer} than the Hawking evaporation time. 

(Note that if we take a relativistic shell with $E\sim M$ instead of $E\sim {P^2\over 2M}$ then we get an even shorter time $t_{dephase}$. Now we would have 
\be
\Delta E\sim \Delta P\gg {1\over R}
\ee
This  gives 
\be
t_{dephase}\ll R \ll MR^2
\ee
where we recall that we are measuring all quantities in planck units, and $M\gg m_{pl}, R\gg l_{pl}$.)

\subsubsection{The effect of phase space volume}

Having obtained a rough picture of how black hole infall may be studied using fuzzball states, let us  consider a toy model which  illustrates in more detail how wavefunctions `spread' during evolution.

\begin{figure}[ht]
\includegraphics[scale=.15]{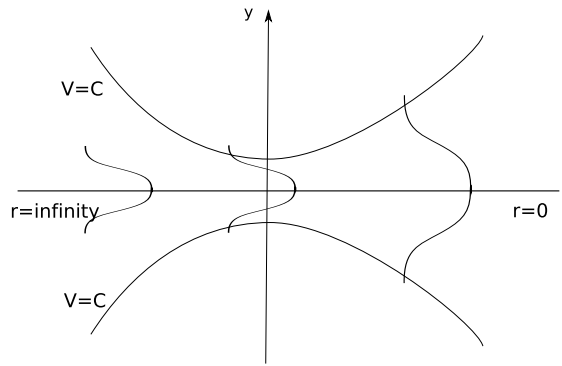}
%
%
\caption{The wavepacket travels in from $r=\infty$ towards $r=0$. The lines of constant potential are sketched; they allow the wavepacket to spread as it reaches $r\rightarrow 0$.}
\label{potential}       
\end{figure}

 In fig.\ref{potential} we sketch a system where a quantum can move along the $r$ direction, from $r=\infty$ to $r=0$. 
If we have only this direction $r$ to move in, the motion of a quantum would be straightforward. But now let us assume that there is another direction $y$ in our space. Let there be a potential
\be
V=\h k(r) y^2
\ee
Let $k(r)$ vanish at large and small $r$ and be high in-between, with the peak at $r=r_0$.

Now let us see what this toy model represents. If $k(r)$ vanishes near $r=0$, then the wavefunction can easily spread over a large range of values of $y$ once the quantum gets close to $r=0$. This represents the fact that there is  a large phase space of fuzzball states (given by the Bekenstein entropy) which can be accessed once an infalling shell comes close enough to the origin. For larger $r$ there are much fewer states for the given energy, while at infinity there are again many states possible because of the large volume of space available. 

First consider a classical particle moving in this $r-y$ space. We can assume $y=0, p_y=0$ consistently, and the particle just reaches the point $r=0, y=0$ at the end of its motion.

Now consider the quantum problem, and start with a wavepacket $e^{-\alpha y^2}$ at large $r$. If $\alpha$ is large enough, the wavepacket will manage to pass through the location of steep potential at $r=r_0$, and emerge into the region at small $r$. But in this region there is no potential limiting the wavefunction in the $y$ direction, so it can spread over the region $-\infty<y<\infty$. 

Thus while the classical solution suggested that the endpoint of the motion is at $r=0, y=0$, the actual wavefunction can spread over all $y$ on reaching $r=0$. This effect becomes more pronounced if we have a large number of transverse directions like $y$. In our actual problem the wavefunction of a collapsing shell can spread over the very large  of $e^{S_{bek}}$ fuzzball states after the shell becomes smaller than a certain size. It is possible that the consequent spreading of the wavefunction invalidates a classical analysis of the motion of the shell. 

\subsubsection{Summary}

Let us summarize the above discussion on the possible dynamics of fuzzballs. A principal feature characterizing black holes is their large entropy. The traditional picture of the hole does not exhibit the microstates required to explain this entropy. If we take the presence of the large number of microstates into account, then the wavefunction of a collapsing shell might spread to a nontrivial extent over this vast phase space of allowed solutions. The resulting dynamics would not correspond to a given quantum moving on a given black hole geometry, but rather lead to a wavefunctional $\psi[g,\phi]$ that is spread over all possible geometries. If this happens then we cannot argue that the light cones of the traditional black hole geometry trap the information of the shell forever and lead to information loss.

\section{Discussion}

So what is the fuzzball proposal and what does it say about the information problem?

Suppose we go to a condensed matter physicist, and tell him about the information paradox. We show him the principles (a), (b) listed in section \ref{introduction}, and tell him that they are reasonable conditions to assume for quantum gravity. We then prove to him that given these conditions, there will {\it have} to be a violation of quantum unitarity. Since the condensed matter person uses quantum theory, he would be very concerned that quantum theory needs a fundamental modification, even though the violation may not be significant in his systems of interest. Indeed, he would probably agree that resolving this paradox should be an important goal of theoretical physics.

Now let us see how the results of the fuzzball program change the situation. The fuzzball proposal does not require new physics, or try to develop abstract principles about what happens in black holes. It simply takes a fully consistent theory of quantum gravity -- string theory -- and explicitly makes examples of microstates of black holes.
All states made so far turn out to be different from the traditional geometry of a black hole: the microstates do not have an `information free horizon'. Thus condition (a) of the Hawking `theorem' breaks down. The fuzzball `conjecture' just says that the microstates not yet constructed would continue to have this feature; thus there should not be two sharply different classes of microstates, one with `information free horizons' and one without. 
 
Given the results of the fuzzball program, what would the condensed matter physicist say? He cannot agree that there is any information {\it paradox}. A paradox is a sharp contradiction that we cannot find a way around. If we can find a way around the paradox for some black hole states, then we cannot argue that there is any sharp contradiction with black holes, even though we have not yet constructed all possible states for all holes. Thus the condensed matter person will simply tell us to go and make other fuzzball states, and come back only if we can show that there are states of black holes that are {\it not} fuzzballs. To summarize, now the `boot is on the other leg'; with the results from the fuzzball program we do not have an information `paradox' unless we can show that the behavior of microstates found so far does not continue in  natural way to the  class of all microstates.

But it is important to note that this does not mean that we understand all there is to know about black holes. For one thing, we still have a lot to learn about the dynamics of black holes. We have conjectured some aspects of dynamics above, and it would be good to check these ideas in concrete detail and to understand what role is played by the large phase space of fuzzball solutions.

In the early days of of the fuzzball program there were some concerns that quantum corrections may destroy the fuzzball nature of 2-charge solutions, and that 3-charge microstates may not be fuzzballs like the 2-charge ones. Possible  quantum corrections were investigated \cite{higher,phase} and no evidence was found that they would be a problem; the magnitude of these corrections was shown to be bounded because of the geometric structure of the fuzzball solution. Large numbers of 3-charge and 4-charge solutions have been made, and now there are also families of nonextremal solutions. For these reasons, perhaps at this point we should accept the hypothesis that the $e^{S_{bek}}$ states of the hole are fuzzballs, and see what this hypothesis tells us about the physics of black holes.

One thing we can do with the fuzzball picture is ask if we can find evidence for various ideas that have been suggested in the past:

\bigskip

(a) It has been suggested  \cite{complementarity}  that the observations made by an infalling observer are given by a description that is `complementary' to the observations made by an observer at infinity. Let us see if we can say anything about this suggestion from our microstate constructions.
In \cite{infall} the infall of a test quantum into the extremal 2-charge system was studied in the CFT picture. It was found that there were {\it three} different logical ways to define time evolution for the quantum: one suited to an infalling quantum,  one to an emerging quantum, and one symmetrical between these two, which may be appropriate for an observer at infinity.  Simple states in one description look very complicated in the other, with the `complication' determined by the entropy of the state. Note  that we do  not have different Hilbert spaces for different observers. Nevertheless, it would be good to see if there is a relation between such effects and notion of complementarity.

(b) Recently it has been suggested \cite{hormalda}   that there is a `future boundary condition' that must be imposed at the black hole singularity. This makes the state at the singularity unique, and forces information to come out in the Hawking radiation. With fuzzballs, we find that states of the hole `swell up' and become big because we need an adequate phase space to hold $e^{S_{bek}}$ states \cite{phase}. Thus with fuzzballs there is a sense in which data cannot be `focused' to a singularity. Perhaps this effect can be interpreted as some kind of a boundary condition at a singularity, and thus a relation found with the idea of a boundary condition at the singularity \cite{hormalda}.

(c) In one of the earliest attempts at resolving the information paradox \cite{thooft}  it was argued that when considering virtual quanta, we should take into account their gravitational backreaction; thus the creation operator for a scalar quantum should be `dressed' with gravity
excitations. For black holes, it was argued that this would lead to large gravitational backreaction from the Hawking radiation quanta, destroying the traditional picture of semiclassical particle production at a low curvature horizon. This proposal has a standard counter-argument:  the gravitational effects of the {\it pair} of produced quanta should cancel out at the horizon, so that the Hawking derivation is not really invalidated. Let us now recall our discussion of  section \ref{spread}, where we have  seen that to follow the effect of an infalling shell we must expand its wavefunction in eigenstates of the {\it total} system (matter+gravity), with energy (\ref{stotal}). So while the argument of \cite{thooft} may not work for the Hawking quanta of a scalar field on a spherically symmetric background,  with the full set of nonperturbative black hole microstates we do find support for the idea that matter states should be studied only with their full gravitational backreaction included.  

(d) There have been studies \cite{shenker} of geodesics in the traditional black hole geometry, where it was found that complex geodesics gave a dominant saddle point describing the correlation of operators in the dual CFT; these correlations were then used as a way of characterization of the singularity. Fuzzballs states are not expected to have such a singularity individually (though the quantum fuzz does get more dense towards the center for a typical state). But when we take an average over fuzzball states, the traditional black hole geometry can appear as a saddle point of the entire sum \cite{phase}, and it would be interesting to see if the complex geodesics emerge naturally to describe expectation values of correlation functions in the ensemble of fuzzball states.

\bigskip

A crucial question now is to extract the essential lessons of the fuzzball program, and see what it tells us about the structure of quantum gravity when we have large amounts of matter crushed at high densities. Clearly, one feature that we have seen is that quantum gravity effects do not extend over a fixed distance like $l_p$; instead this distance increases with the number of quanta involved in the black hole bound state. What does this tell us about Cosmology, where we also have large amounts of matter at high densities? In \cite{cosmology} the state of the early Universe was modeled after the states that give the entropy of black holes, and the resulting evolution was studied. The Universe did not inflate. But  the nonlocal correlations in the quantum bound state extended right across the Universe.  So we might have a different possible resolution of the `horizon problem':  the Universe is homogeneous because of quantum nonlocal effects at very early times.  There are many other questions that we need to answer however before we can get a proper understanding of the early Universe. What determines the initial state? If this state to be determined from first principles, or to be randomly chosen from a given set? We do not know how to address such questions yet, but knowing the nature of black hole microstates should be a start in understanding the dense matter that must almost certainly be a feature of the far past.

\section*{Acknowledgments}

I would like to thank Steve Avery, Borun Chowdhury,  Sumit Das, Stefano Giusto and  Oleg Lunin  for many helpful comments. 
This work was supported in part by DOE grant DE-FG02-91ER-40690.

\end{document}